# TRUST-BASED WINNOW LINEAR MULTIPLICATIVE CLASSIFICATION FOR SECURE MULTIPATH ROUTING IN MANET

**Mrs. A.S.Narmadha[1],**

**Research Scholar, Jai Shriram Engineering College, Tirupur**

**Dr.S.Maheswari[2],**

**AP (Sr.Gr)/EEE, Kongu Engineering College, Erode**

## ABSTRACT

Multipath routing in Mobile Ad Hoc Network (MANET) plays a significant concern for secured data transmission by avoiding the attack nodes in the network. Few research works have been developed for secured multipath routing in MANET. Data security level during the process of multipath routing was not sufficient. Besides, the amount of time needed to perform secure multipath routing was more. In order to overcome such limitations, a Winnow Trust based Multipath Route Discovery (WT-MRD) Mechanism is proposed. The designed WT-MRD Mechanism constructs multiple paths from source to destination with higher security and lesser time. Initially, the trust value of each node is calculated based on the node cooperative count, data packet forwarding rate and packet drop rate by Neighbor Node-based Trust Calculation (NN-TC) Model. After calculating the trust value, the nodes are classified as normal or malicious by using Winnow Linear Multiplicative Classification (WLMC) Algorithm. With the help of normal nodes, the WT-MRD Mechanism finds multipath from source to destination by sending a two control message RREQ and RREP. Source node transmits a route request RREQ to the neighboring node for constructing the multiple route paths. After receiving the RREQ message, the neighboring node maintains the route table where the source information and next hop information are present. Then Route Reply (RREP) messages are sent from neighboring node to source node. By this way, multiple route paths from source to destination are constructed with a higher security level. Simulation evaluation of WT-MRD Mechanism is conducted on factors such as attack detection rate, attack detection time and data security level and delay with respect to a number of mobile nodes and data packets.

**Keywords:** Cooperative Count, Data Packet Dropped Rate, Data Packet Forwarded Rate, Multipath Routing, Security, Trust Value, Winnow Linear Multiplicative Classification.

## 1. INTRODUCTION

Recently, MANET gets greater attention due to their self-configuration and self-maintenance characteristic. Routing in MANET is a vital task for broadcasting the data packets from source to destination. Multipath routing is the process of transmitting the data from the source node to the destination node through multiple paths at the same time. MANET is affected by many security attacks. Therefore, security is a significant problem to be resolved in MANET for preserving communication between nodes. Some of the research works have been designed for secured multiple routing in MANET. But in many existing works, data security level during the multiple routing was not improved. For improving the communication security in MANET, the attacks are to be detected during the routing process. Hence, a novel WT-MRD Mechanism is introduced in this research work.

Trust enhanced cluster-based multipath routing (TECM-OLSR) algorithm was introduced in [1] with the help of Particle swarm optimization. But, the data security level was not enough. Secured and QoS based energy-aware multipath routing was performed in [2] with the application of Particle swarm optimization-genetic search algorithm (PSOGSA). However, packet delivery ratio and delay time were not at the required level.

A new method was introduced in [3] for node authentication when a new node joins into the network before initiating route discovery process in MANET. But, end to end delay was more. In [4], Flooding Factor-based Framework for Trust Management (F3TM) was carried out in MANETs. But, F3TM was taken a large amount of time to calculate the trust value.

A secure and energy-efficient stochastic multipath routing protocol was introduced in [5] depending on the Markov chain for MANETs. Though the energy consumption was reduced, the security level was not improved. The standard ad hoc on-demand multipath distance vector protocol was extended as the base routing protocol in [6]. However, the Dolphin Echolocation

Algorithm failed to identify the malicious node through a classification process which reduced the communication security level.

An anonymous multipath routing protocol was presented in [7] by using secret sharing to obtain higher security in MANET. But, the data loss rate using this protocol was more. In [8], an efficient and stable multipath routing was carried out with congestion awareness. However, secure multipath routing was not obtained.

The Multipath Dijkstra Algorithm was applied in [9] to find multiple paths and thereby improving the quality of service in MANET. But, data security level was lower. A novel technique was designed in [10] using fitness function to discover the optimal path from the source node to the destination node and to lessen the energy consumption in multipath routing. However, packet delivery was not improved.

In order to address the above said existing drawbacks in MANET, WT-MRD Mechanism is developed in this research work. The key contribution of WT-MRD Mechanism is presented in below,

- ❖ To enhance the security level of multipath routing in MANET when compared to state-of-the-art works, Winnow Trust based Multipath Route Discovery (WT-MRD) Mechanism is developed. On the contrary to conventional works, WT-MRD Mechanism is designed by combining Neighbor Node-based Trust Calculation (NN-TC) Model and Winnow Linear Multiplicative Classification (WLMC) Algorithm.

- ❖ To improve the attack detection performance with a minimal time complexity when compared to conventional works, NN-TC Model and WLMC Algorithm are employed in WT-MRD Mechanism. On the contrary to existing works, NN-TC Model computes trust value for each mobile node based on the factors i.e. node cooperative count, data packet forwarding rate and packet drop rate for accurately detecting malicious node in a network.

The rest of the paper is prepared as follows. Section 2 presents the detailed process of proposed WT-MRD Mechanism with the help of the architecture diagram. In section 3, a simulation setting of WT-MRD Mechanism is presented. Section 4 provides the comparative results of proposed Mechanism. Section 5 portrays the literature survey. At last, the conclusion of the paper is depicted in Section 6.

## 2. WINNOW TRUST-BASED MULTIPATH ROUTE DISCOVERY

MANET is a self-organized system which includes many wireless mobile nodes. All nodes operate as both communicators and routers. Due to multi-hop routing and lack of centralized administration in an open environment, MANET is susceptible to malicious attacks nodes. Therefore, secured multipath routing is a considerable problem in MANET to achieve higher packet delivery ratio and a minimal end to end delay. In conventional works, many intrusion detection techniques were developed for performing secure multipath routing in MANET. However, higher security was not obtained during the routing process which impacts packet delivery ratio and delay in MANET. In order to overcome such limitations, Winnow Trust based Multipath Route Discovery (WT-MRD) Mechanism is introduced in this research work. The architecture diagram of WT-MRD Mechanism is presented in below Figure 1.

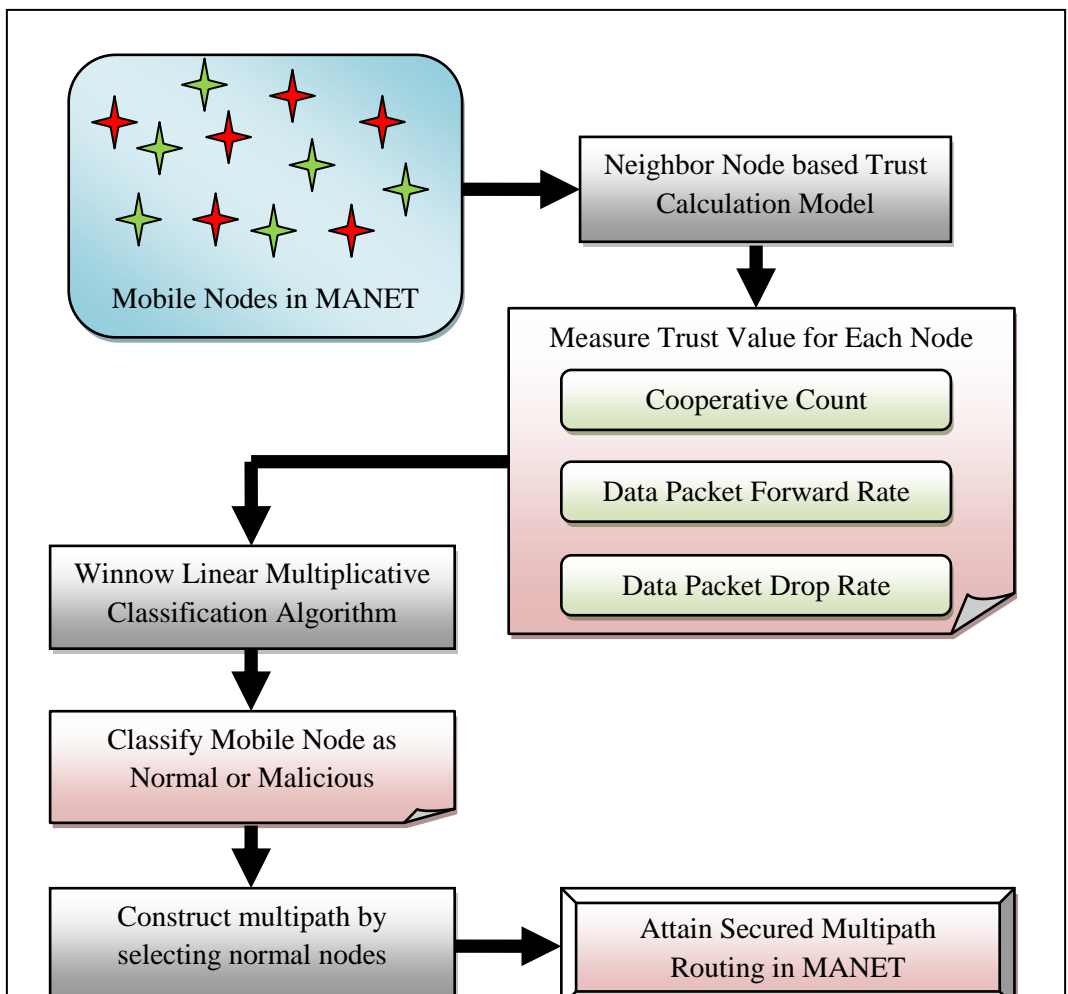

**Figure 1 WT-MRD Mechanism for Secured Multipath Routing in MANET**

Figure 1 depicts the overall processes of proposed WT-MRD Mechanism to get enhanced security level while carried outing the multipath routing in MANET. As demonstrated in the above figure, WT-MRD Mechanism at first randomly initializes a number of mobile nodes in a network. Followed by, WT-MRD Mechanism applies Neighbor Node-based Trust Calculation (NN-TC) Model where it calculates trust value for all mobile nodes by considering the factors such as node cooperative count, data packet forwarding rate and packet drop rate in a network. Based on the estimated trust values, then WT-MRD Mechanism classifies each mobile node as normal or malicious with the application of Winnow Linear Multiplicative Classification (WLMC) Algorithm. From that, WT-MRD Mechanism accurately finds the malicious node in MANET with higher accuracy and minimal time. By considering normal nodes, finally, WT-MRD Mechanism builds multipath between the source and destination via broadcasting two control message RREQ and RREP. Therefore, WT-MRD Mechanism enhances the security level of multipath routing in MANET when compared to state-of-the-art works.

**2.1 Neighbor Node-based Trust Calculation (NN-TC) Model**

In WT-MRD Mechanism, Neighbor Node-based Trust Calculation (NN-TC) Model is designed in order to significantly determine the trust value for each mobile node in MANET based on the factors such as node cooperative count, data packet forwarding rate and packet drop rate. Owing to the arbitrary movement of the mobile nodes in MANET, cooperative communication is very important. In NN-TC Model, the cooperativeness of mobile nodes is estimated based on the behavior of that node when connecting with other nodes. Therefore, node cooperative count determines the connections between mobile nodes change over time. Hence, each mobile node in MANET is responsible for managing the connections to all its neighbors.

Whenever the neighboring nodes connections are protected, the end-to-end connectivity is then assured to perform reliable data delivery in MANET.

Let us consider '$\mu_i$' and '$\mu_j$' are two random mobile nodes in the network. The mobile node '$\mu_i$' transmit data packet '$\rho_i$' to mobile node '$\mu_j$' at an arbitrary time. Before transmission, the connection between the two nodes is created by sending the two control messages namely RREQ and RREP. The mobile node '$\mu_i$' broadcast RREQ to mobile node '$\mu_j$' using below mathematical expression,

$$\mu_i \ (RREQ) \rightarrow \mu_j \qquad (1)$$

Then, the mobile node '$\mu_j$' gets the RREQ message and send reply messages ($RREP$) to mobile node '$\mu_i$' using below mathematical representation,

$$\mu_j(RREP) \rightarrow \mu_i \qquad (2)$$

When mobile node '$\mu_j$' broadcasts a reply message to the mobile node '$\mu_i$' in a network, $\mu_i$ is connected to the '$\mu_j$' at the time for efficient data transmission. If the mobile node '$\mu_i$' does not get any reply message, mobile nodes '$\mu_i$' and '$\mu_j$' are not connected. Based on this connections establishment, NN-TC Model estimates cooperative count '$\alpha_{\mu_i}$' for each mobile node in MANET with a minimal amount of time.

In NN-TC Model, the data packets forwarded rate is determined as the percentages of the number of data packets forwarded by the mobile node '$\mu_i$' over the total number of data packets distributed to mobile node '$\mu_i$'. From that, the data packets forwarded rate '$\beta_{\mu_i}$' is mathematically calculated using below formulation,

$$\beta_{\mu_i} = \frac{DPF(\mu_i)}{m_i} \qquad (3)$$

From the above mathematical expression (3), '$DPF(\mu_i)$' point outs the number of data packets forwarded by the node '$\mu_i$' whereas '$m_i$' indicates the number of data packets received by the mobile node '$\mu_i$' in a MANET.

In NN-TC Model, the data packets dropped rate is calculated as percentages of the packets that were dropped over the total number of data packets transmitted to the mobile node $\mu_i$. Thus, the data packets dropped rate '$\gamma_{\mu_i}$' is mathematically evaluated using below expression,

$$\gamma_{\mu_i} = \frac{DPD(\mu_i)}{mi} \qquad (4)$$

From the above mathematical equation (4), '$DPD(\mu_i)$' denotes the number of data packets lost by the mobile node '$\mu_i$' and '$m_i$' symbolizes the number of data packets get by the mobile node '$\mu_i$' in a network.

Based on the determined node cooperative count '$\alpha_{\mu_i}$' data packets forwarded rate '$\beta_{\mu_i}$' and data packets dropped rate '$\gamma_{\mu_i}$', finally NN-TC Model measures the trust value for each mobile node '$\tau_{\mu_i}$' using below mathematical formula,

$$\tau_{\mu_i} = \alpha_{\mu_i} + \beta_{\mu_i} + \gamma_{\mu_i} \qquad (5)$$

By using the above mathematical representation (5), NN-TC Model calculates trust value '$\tau_{\mu_i}$' for each mobile node in order to discovers the normal and malicious nodes in network through classification.

The algorithmic processes of Neighbor Node-based Trust Calculation (NN-TC) Model is explained in below,

---

// **Neighbor Node-based Trust Calculation Algorithm**
**Input:** Number of Mobile Nodes '$\mu_i = \mu_1, \mu_2, \mu_3 \ldots \mu_n$'; Number of Data Packets '$\rho_i =$

$\rho_1, \rho_2, \rho_3 \ldots \rho_N$'

**Output:** Get Trust Value for all mobile nodes

**Step 1:** Begin

**Step 2:**     **For** each mobile node '$\mu_i$'

**Step 3:**        Calculate cooperative count '$\alpha_{\mu_i}$' using (1) and (2)

**Step 4:**        Determine data packets forwarded rate '$\beta_{\mu_i}$' using (3)

**Step 5:**        Measure data packets dropped rate '$\gamma_{\mu_i}$' using (4)

**Step 6:**        Estimate trust value '$\tau_{\mu_i}$' using (5)

**Step 7:**     **End For**

**Step 8:** End

**Algorithm 1 Neighbor Node-based Trust Calculation**

Algorithm 1 explains step by step process of NN-TC Model. By using the above algorithmic processes, NN-TC Model computes trust value for all the mobile nodes for separating normal and malicious nodes in MANET and thereby obtaining secure multipath routing.

**2.2 Winnow Linear Multiplicative Classification Algorithm**

In WT-MRD Mechanism, Winnow Linear Multiplicative Classification (WLMC) Algorithm is a machine learning technique that learns a linear classifier for categorizing mobile nodes as normal or malicious based on their trust value. The WLMC is very similar to the perceptron algorithm. The perceptron algorithm employs an additive weight-update scheme. On the contrary to this, WLMC utilizes a multiplicative scheme that allows it to carry out better classification when many dimensions are irrelevant (therefore it's called winnow). The WLMC is a simple algorithm that scales well while considering the large size of MANET. During the classification process, WLMC is shown a sequence of positive (i.e. normal mobile nodes) and negative examples (i.e. malicious mobile nodes). From these, WLMC learns a decision hyperplane that used to classify each mobile node as positive (i.e. normal mobile nodes) or negative (i.e. malicious mobile nodes) with higher accuracy and minimal time complexity.

The WLMC algorithm initially takes a number of mobile nodes '$\mu_1, \mu_2, .., \mu_n$' in MANET as input where '$n$' signifies a total number of nodes considered in a network. The instance space is '$\mu = \{0, 1\}$' i.e. each input mobile node is described as a set of Boolean-valued features. Then WLMC define non-negative weights '$\omega_i$' for each mobile node '$i\epsilon\{1, ..., n\}$' which are initially set to '1', one weight for each mobile node. Followed by, WLMC algorithm applies the typical prediction rule to classify the each mobile node according to trust value using below mathematical representation,

$$z^* = \begin{cases} if\ (\omega^T \mu_i > \theta), then\ z^* = +1 \\ otherwise \qquad\qquad\quad z^* = -1 \end{cases} \quad (6)$$

From the above equation (6), '$\theta$' denotes a real number (i.e. threshold) whereas '$z^*$' indicates the classified output of WLMC algorithm (predicted output). Together with the weight, the threshold determines a separating hyperplane in the instance space for accurately categorizing mobile node in MANET as normal or malicious. In WLMC algorithm, good bounds are attained if '$\theta = n/2$'.

For each obtained classification result, then WLMC algorithm applies the following update rule:

- ✓ If an input mobile node is correctly classified, do nothing.
- ✓ If an input mobile node is predicted incorrectly and the correct result was '+1', then the WLMC algorithm updates each weight multiplied by '2' i.e. '$\omega_i \leftarrow 2\omega_i$' only for those mobile node features '$\mu_i$' that are '1'.
- ✓ If an input mobile node is predicted incorrectly and the correct result was '-1', then the WLMC algorithm updates each weight divided by '2' '$\omega_i \leftarrow \omega_i/2$' only for those mobile node features '$\mu_i$' that is '-1'.

By using the above processes, the WLMC algorithm significantly classifies the nodes in MANET as a positive node or negative node as depicted in below figure 2 based on estimated trust values.

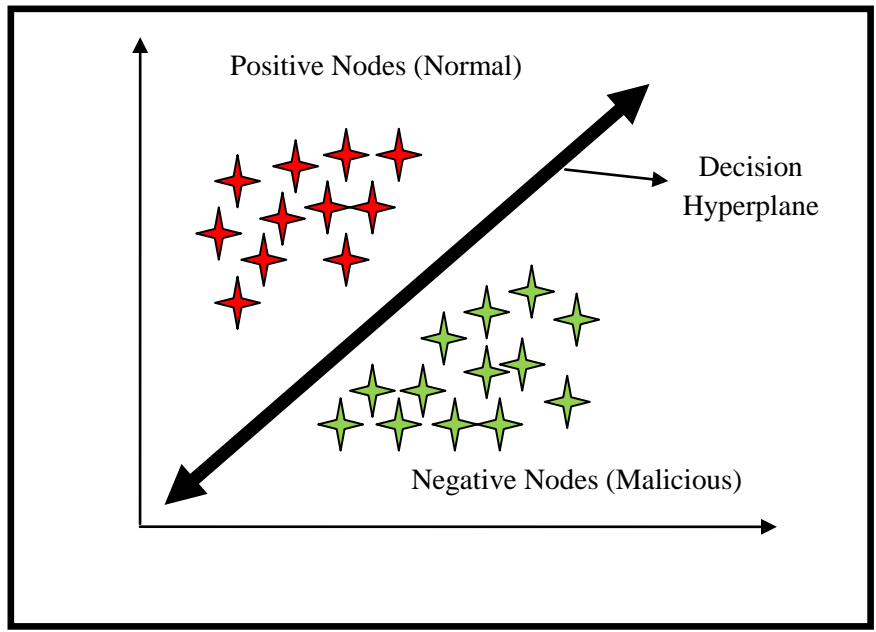

**Figure 2 Examples for WLMC algorithm**

The algorithmic processes of WLMC algorithm are presented in below,

---

// **Winnow Linear Multiplicative Classification Algorithm**

**Input:** Number of Mobile Nodes '$\mu_i = \mu_1, \mu_2, \mu_3 \ldots \mu_n$'; Actual Output '$z$'; Predicted Output '$z^*$'; weight '$\omega$'; Source Node 'S'; Destination node 'D'; Number of Data Packets '$\rho_i = \rho_1, \rho_2, \rho_3 \ldots \rho_N$'

**Output:** Achieve higher attack detection rate

**Step 1:** Begin

**Step 2:**    **For** each mobile node '$\mu_i$'

**Step 3:**       Initialize weight $\omega = 1$

**Step 4:** Apply prediction rule to get classification result using (6)

**Step 5:** **If** $(z=+1)$ and $(z^* = -1)$ then

**Step 6:** Update each weight $\omega_i \leftarrow 2\omega_i$ only for node features '$\mu_i$' with '1'

**Step 7:** **Else If** $(z=-1)$ and $(z^* = +1)$ then

**Step 8:** Update each weight $\omega_i \leftarrow \omega_i/2$ only for node features '$\mu_i$' with '1'

**Step 9:** **End If**

**Step 10:** **If $(z == +1)$, then**

**Step 11:** '$\mu_i$' is a normal node

**Step 12:** **Else If $(z == -1)$, then**

**Step 13:** '$\mu_i$' is malicious node

**Step 14:** **End If**

**Step 15:** **End For**

**Step 16:** **For** each normal node

**Step 17:** '$S$' sent RREQ message to its all neighboring node

**Step 18:** Neighboring node sent RREP messages to 'S'

**Step 19:** Keep routing table

**Step 20:** Create secured multiple paths between 'S' and 'D' to route data packets '$\rho_i$'

**Step 21:** **End For**

**Step 22:** End

**Algorithm 2 Winnow Linear Multiplicative Classification**

Algorithm 2 shows the step by step process of WLMC algorithm to efficiently classify all the mobile nodes in MANET into different classes (positive nodes or negative nodes) with higher accuracy. Thus, WT-MRD Mechanism considers only normal mobile nodes to establish multipath between source and destination by sending two control message RREQ and RREP. Source node broadcasts a route request RREQ to the neighboring node for creating the multiple route path in a network. After getting the RREQ message, the neighboring node maintains the route table where the source information and next hop information are present. Then Route Reply (RREP) messages are sent from neighboring node to source node. In this way, multiple route paths from source to destination are constructed with a higher security level for routing data packets in MANET as compared to other conventional works. As a result, WT-MRD Mechanism increases the performance of multipath routing with higher packet delivery ratio and a minimal end to end delay as compared to state-of-the-art works.

## 3. SIMULATION SETTINGS

In order to measure the performance of the proposed, WT-MRD Mechanism is implemented in NS2.34 simulator by considering a different number of mobile nodes in a square area of $A^2$ (1200 m * 1200 m). The simulation parameters for conducting the simulation process in WT-MRD Mechanism shown in below Table 1.

**Table 1 Simulation Parameters**

| Simulation parameter | Value |
|---|---|
| Simulator | NS2 .34 |
| Protocol | AODV |
| Number of mobile nodes | 50,100,150,200,250,300,350,400,450,500 |
| Simulation time | 250 sec |
| Mobility model | Random Way Point model |
| Nodes speed | 0-20m/s |
| Network area | 1200m * 1200m |
| Data packets | 10,20,30,40,50,60,70,80,90,100 |
| Number of runs | 10 |
| Traffic type | CBR |

The performance of WT-MRD Mechanism is estimated in terms of attack detection rate, attack detection time and data security level, and delay.

## 4. RESULTS

In this section, the simulation result of WT-MRD Mechanism is compared with two state-of-the-art works namely TECM-OLSR [1] and PSOGSA [2]. The effectiveness of WT-MRD Mechanism is determined with the aid of below tables and graphs.

**4.1 Attack Detection Rate**

In WT-MRD Mechanism, Attack Detection Rate '$(ADR)$' calculates the ratio of number of mobile nodes that are exactly detected as normal or malicious to the total number mobile nodes considered for simulation process. From that, the attack detection rate is mathematically determined as,

$$ADR = \frac{N_{\mu CD}}{\mu_n} * 100 \qquad (7)$$

From the above formula (7), '$\mu_n$' point outs a total number of mobile nodes whereas '$N_{\mu CD}$' indicates a number of mobile nodes that are correctly detected. The attack detection rate is estimated in terms of percentages (%).

**Sample Mathematical Calculations:**

- ❖ **Proposed WT-MRD:** number of mobile nodes correctly discovered as normal or malicious is 47 and the total number of the mobile node is 50. Then, the attack detection rate is obtained as,

$$ADR = \frac{47}{50} * 100 = 94\ \%$$

- ❖ **Existing TECM-OLSR:** number of mobile nodes exactly identified as normal or malicious is 39 and the total number of mobile nodes is 50. Then, the attack detection rate is estimated as,

$$ADR = \frac{39}{50} * 100 = 78\ \%$$

- ❖ **Existing PSOGSA:** number of mobile nodes precisely detected as normal or malicious is 36 and the total number of the mobile node is 50. Then, the attack detection rate is evaluated as,

$$ADR = \frac{36}{50} * 100 = 72\ \%$$

The simulation result analysis of attack detection rate during multipath routing is demonstrated in below Table 2.

**Table 2 Tabulation for Attack Detection Rate**

| Number of mobile nodes | Attack Detection Rate (%) | | |
|---|---|---|---|
| | WT-MRD | TECM-OLSR | PSOGSA |
| 50 | 94 | 78 | 72 |
| 100 | 92 | 73 | 70 |
| 150 | 93 | 81 | 79 |
| 200 | 95 | 83 | 82 |
| 250 | 93 | 79 | 78 |
| 300 | 95 | 84 | 83 |
| 350 | 96 | 81 | 80 |
| 400 | 98 | 89 | 88 |
| 450 | 96 | 84 | 83 |
| 500 | 95 | 86 | 85 |

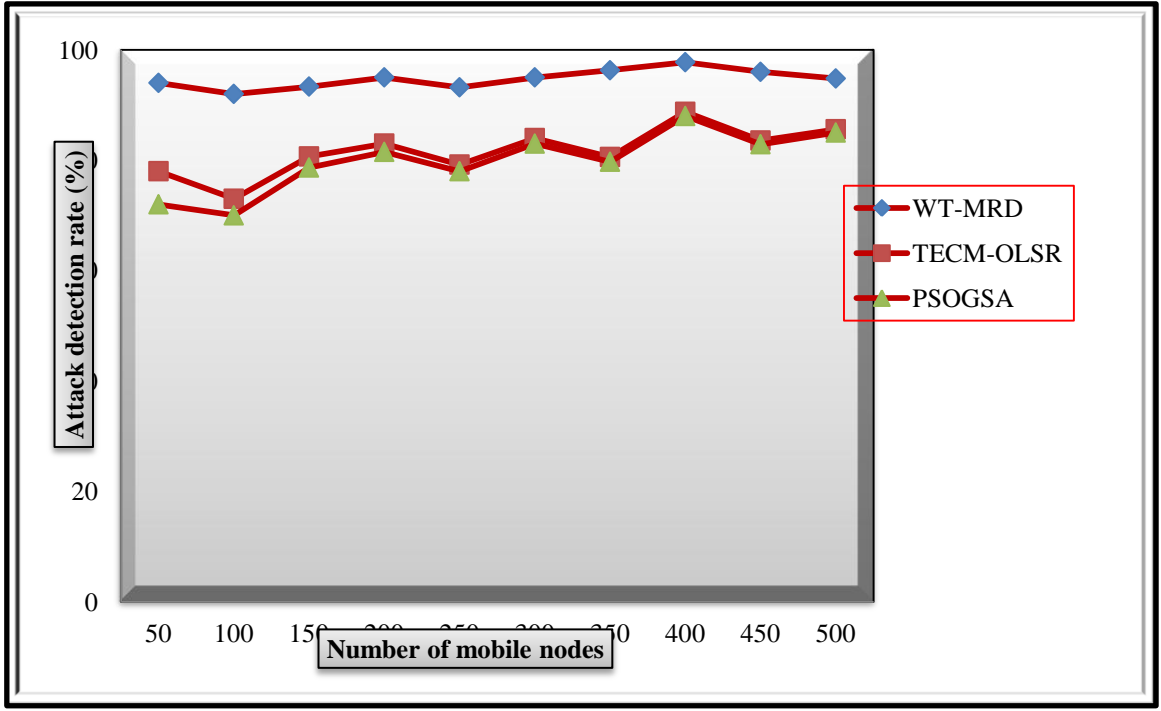

**Figure 3 Comparative Result Analysis of Attack Detection Rate Versus the number of Mobile Nodes**

Figure 3 shows the simulation measure of attack detection rate based on a diverse number of mobile nodes using three methods namely proposed WT-MRD and existing TECM-OLSR [1] and PSOGSA [2]. As presented in the above graph, the proposed WT-MRD mechanism attains higher attack detection rate for finding malicious attack node in MANET as compared to conventional TECM-OLSR [1] and PSOGSA [2]. This is owing to the application of Neighbor Node-based Trust Calculation (NN-TC) Model and Winnow Linear Multiplicative Classification (WLMC) Algorithm. By using the concepts of Neighbor Node-based Trust Calculation (NN-TC) Model, proposed WT-MRD mechanism computes trust value of mobile nodes by considering node cooperative count, data packet forwarding rate and packet drop rate on the contrary to state-of-the-art works.

This helps for Winnow Linear Multiplicative Classification (WLMC) Algorithm in proposed WT-MRD mechanism to precisely discover the malicious attack nodes in MANET with higher accuracy. Hence, the proposed WT-MRD mechanism increases the ratio of number of mobile nodes that are exactly detected as normal or malicious as compared to other conventional works [1] and [2]. Therefore, the proposed WT-MRD mechanism improves the attack detection rate in MANET by 16 % and 19 % as compared to TECM-OLSR [1] and PSOGSA [2] respectively.

**4.2 Attack Detection Time**

In WT-MRD Mechanism, Attack detection time (ADT) calculates the amount of time required to detect malicious attack nodes in MANET through classification. The attack detection time is measured mathematically as,

$$ADT = \mu_n * t(DS) \qquad (8)$$

From the above equation (8), '$t(DS)$' indicates the time taken for classifying single node whereas '$\mu_n$' refers to a total number of mobile nodes. The attack detection time is computed in terms of milliseconds (ms).

**Sample Mathematical Calculations:**

- ➢ **Proposed WT-MRD:** time consumed to classify single node is 0.4 ms and a total number of mobile nodes is 50. Then, the attack detection time is determined as,

$$ADT = 50 * 0.4 = 20 \, ms$$

- ➢ **Existing TECM-OLSR:** time utilized to classify single node is 0.58 ms and a total number of mobile nodes is 50. Then, the attack detection time is measured as,

$$ADT = 50 * 0.58 = 29 \, ms$$

- ➢ **Existing PSOGSA:** time needed to classify single node is 0.7 ms and a total number of mobile nodes is 50. Then, the attack detection time is obtained as,

$$ADT = 50 * 0.7 = 35\ ms$$

The tabulation result analysis of attack detection time involved during multipath routing is depicted in below Table 3.

**Table 3 Tabulation for Attack Detection Time**

| Number of mobile nodes | Attack Detection Time (ms) | | |
|---|---|---|---|
| | WT-MRD | TECM-OLSR | PSOGSA |
| 50 | 20 | 29 | 35 |
| 100 | 26 | 38 | 42 |
| 150 | 32 | 42 | 48 |
| 200 | 44 | 48 | 52 |
| 250 | 43 | 53 | 60 |
| 300 | 45 | 48 | 63 |
| 350 | 42 | 60 | 67 |
| 400 | 50 | 56 | 68 |
| 450 | 54 | 59 | 72 |
| 500 | 56 | 63 | 75 |

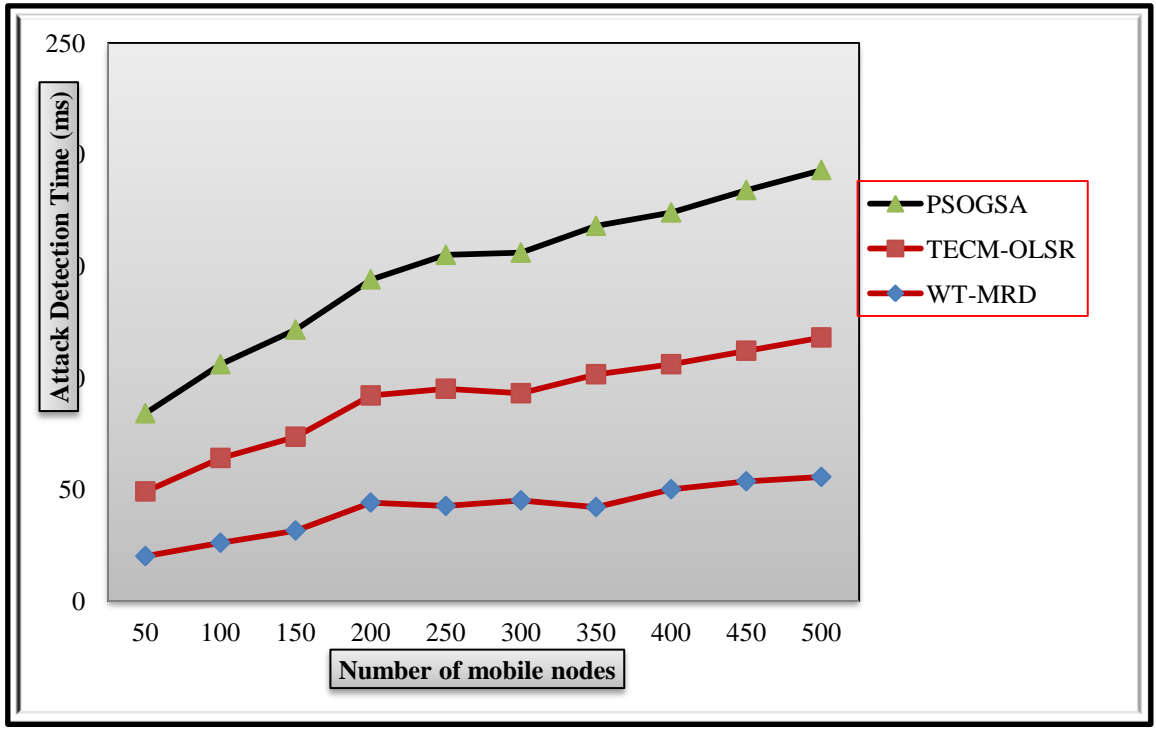

**Figure 4 Comparative Result Analysis of Attack Detection Time Versus the number of Mobile Nodes**

Figure 4 portrays the impact of attack detection time with respect to a varied number of mobile nodes using three methods namely proposed WT-MRD and existing TECM-OLSR [1] and PSOGSA [2]. As shown in the above graphical diagram, proposed WT-MRD mechanism gets minimal attack detection time in MANET as compared to conventional TECM-OLSR [1] and PSOGSA [2]. This is because of the application of Neighbor Node-based Trust Calculation (NN-TC) Model and Winnow Linear Multiplicative Classification (WLMC) Algorithm. With the support of Neighbor Node-based Trust Calculation (NN-TC) Model, proposed WT-MRD mechanism calculates trust value for all mobile nodes according to their node cooperative count, data packet forwarding rate and packet drop rate on the contrary to existing works.

By using the evaluated trust values, then Winnow Linear Multiplicative Classification (WLMC) Algorithm perfectly classify the mobile node as normal or malicious with a lower amount of time utilization. Therefore, the proposed WT-MRD mechanism minimizes the amount of time required to detect malicious attack nodes in MANET as compared to other conventional works [1] and [2]. As a result, the proposed WT-MRD mechanism decreases the attack detection time in MANET by 18 % and 30 % as compared to TECM-OLSR [1] and PSOGSA [2] respectively.

**4.3 Data Security Level**

In WT-MRD Mechanism, Data Security Level '$DSL$' is measured in terms of packet delivery ratio. Thus, data security level is determined as a ratio number of data packets successfully received to destination from source node in MANET. Accordingly, data security level is mathematically calculated as,

$$DSL = \frac{n_{\rho SD}}{\rho_N} * 100 \qquad (9)$$

From the above mathematical formula (9), '$\rho_N$' represent a total number of data packets whereas '$n_{\rho SD}$' indicates number of data packets successfully delivered at destination. The data security level is measured in terms of percentage (%).

**Sample Mathematical Calculation:**

➢ **Proposed WT-MRD**: Number of data packet successfully reached at the destination is 9, and the total number of data packets is 10. Then, the data security level is determined as,

$$DSL = \frac{9}{10} * 100 = 90\%$$

➢ **Existing TECM-OLSR:** Number of data packet successfully attained at the destination is 8, and the total number of data packets is 10. Then, the data security level is computed as,

$$DSL = \frac{8}{10} * 100 = 80\%$$

➢ **Existing PSOGSA:** Number of data packet successfully arrived at destination is 7, and the total number of data packets is 10. Then, the data security level is obtained as,

$$DSL = \frac{7}{10} * 100 = 70\%$$

The performance result analysis of data security level during multipath routing is presented in below Table 4.

**Table 4 Tabulation for Data Security Level**

| Number of data packets | Data Security Level (%) | | |
|---|---|---|---|
| | **WT-MRD** | **TECM-OLSR** | **PSOGSA** |
| 10 | 90 | 80 | 70 |
| 20 | 90 | 75 | 70 |
| 30 | 93 | 83 | 73 |
| 40 | 95 | 85 | 70 |
| 50 | 94 | 76 | 66 |
| 60 | 93 | 87 | 67 |
| 70 | 94 | 89 | 76 |
| 80 | 96 | 85 | 76 |
| 90 | 93 | 82 | 74 |
| 100 | 97 | 84 | 78 |

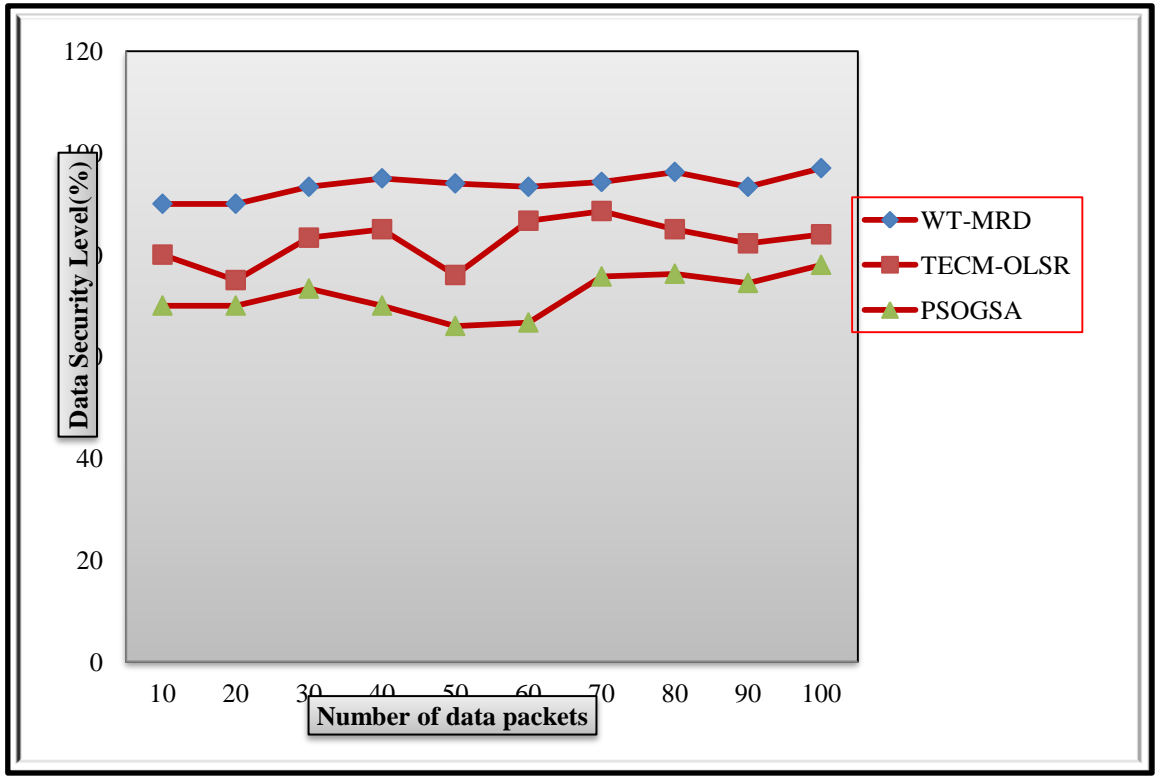

**Figure 5 Comparative Result Analysis of Data Security Level Versus the number of Data Packets**

Figure 5 illustrates the comparative result analysis of data security level along with a different number of data packets using three methods namely proposed WT-MRD and existing TECM-OLSR [1] and PSOGSA [2]. As depicted in the above graphical figure, proposed WT-MRD mechanism achieves higher data security level in MANET when compared to conventional TECM-OLSR [1] and PSOGSA [2]. This is due to the application of Neighbor Node-based Trust Calculation (NN-TC) Model and Winnow Linear Multiplicative Classification (WLMC) Algorithm on the contrary to state-of-the-art works.

By using the algorithmic process of NN-TC Model and WLMC, proposed WT-MRD mechanism chooses a nearby normal node to create secure multipath between source and destination in MANET. From that, the proposed WT-MRD mechanism performs reliable data transmission in MANET as compared to conventional works. Hence, the proposed WT-MRD mechanism improves the ratio number of data packets successfully received to the destination from the source node in MANET as compared to other existing works [1] and [2]. Accordingly, the proposed WT-MRD mechanism enhances the data security level of multipath routing in MANET by 13 % and 30 % as compared to TECM-OLSR [1] and PSOGSA [2] respectively.

**4.4 Delay**

In WT-MRD Mechanism, delay '$D$' measure the distinction between the actual arrival time and expected arrival time of data packets at the destination node in MANET. From that, delay is mathematically determined as,

$$D = (\rho_{AAT} - \rho_{EAT}) \qquad (10)$$

From the above mathematical expression (10), '$\rho_{AAT}$' signifies actual arrival time of a data packets and '$\rho_{EAT}$' denotes an expected arrival time of data packets. The delay is calculated in terms of milliseconds (ms).

**Sample Mathematical Calculation:**

- ➢ **Proposed WT-MRD**: Actual arrival time of data packets is 32 ms and the expected arrival time is 25 ms. Then the delay is measured as,

$$D = 32ms - 25ms = 7ms$$

- ➢ **Existing TECM-OLSR:** Actual arrival time of data packets is 37 ms and the expected arrival time is 25 ms. Then the delay is calculated as,

$$D = 37ms - 25ms = 12ms$$

- ➢ **Existing PSOGSA:** Actual arrival time of data packets is 42 ms and the expected arrival time is 25ms. Then the delay is computed as,

$$D = 42ms - 25ms = 17ms$$

The experimental result analysis of delay involved during the secured multipath routing is shown in below Table 5.

**Table 5 Tabulation for Delay**

| Number of data packets | Delay (ms) | | |
|---|---|---|---|
| | WT-MRD | TECM-OLSR | PSOGSA |
| 10 | 7 | 12 | 17 |
| 20 | 10 | 17 | 24 |
| 30 | 11 | 22 | 30 |
| 40 | 12 | 24 | 32 |
| 50 | 16 | 27 | 30 |
| 60 | 14 | 26 | 34 |
| 70 | 15 | 30 | 37 |
| 80 | 17 | 32 | 36 |
| 90 | 18 | 30 | 39 |
| 100 | 16 | 34 | 38 |

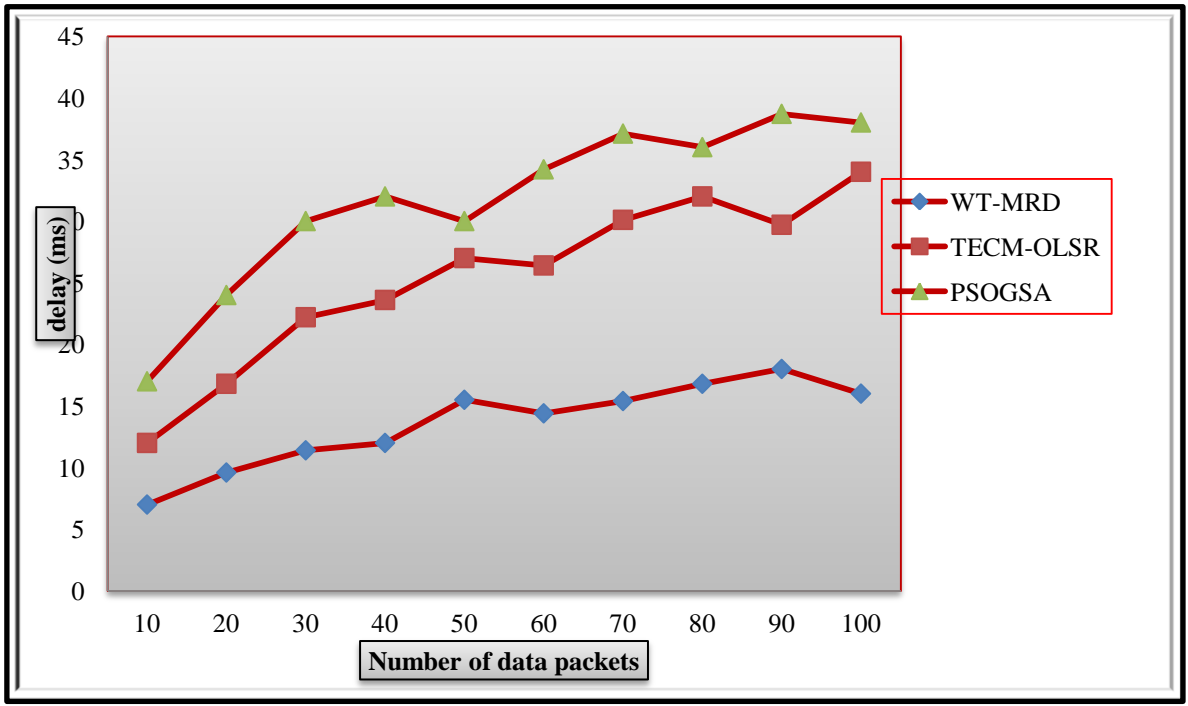

**Figure 6 Comparative Result Analysis of Delay versus Number of Data Packets**

Figure 6 presents performance result analysis of delay with respect to a various number of data packets using three methods namely proposed WT-MRD and existing TECM-OLSR [1] and PSOGSA [2]. As shown in the above graphical diagram, proposed WT-MRD mechanism obtains a lower amount of delay when carried outing multipath routing process in MANET when compared to conventional TECM-OLSR [1] and PSOGSA [2]. This is because of the application of NN-TC Model and WLMC Algorithm on the contrary to conventional works.

With the support of both the NN-TC Model and WLMC algorithmic steps, proposed WT-MRD mechanism performs secure multipath routing in MANET. Thus, the proposed WT-MRD mechanism attains reliable data delivery with a minimal amount of time complexity in MANET as compared to state-of-the-art works. For this reason, proposed WT-MRD mechanism minimizes the difference between the actual arrival time and expected arrival time of data packets at destination node in MANET as compared to other existing works [1] and [2]. Therefore, the proposed WT-MRD mechanism minimizes the delay involved during the secured multipath routing in MANET by 46 % and 57 % as compared to TECM-OLSR [1] and PSOGSA [2] respectively.

## 5. LITERATURE SURVEY

Two multi-paths routing protocols were applied in [11] to get better quality of service for video streaming in MANET. A game-theoretic framework was presented in [12] for stochastic multipath routing in a network.

A trust-based reactive multipath routing protocol was used in [13] to learn multiple loop-free paths in MANET. Expected residual lifetime based ad hoc on-demand multipath routing protocol (ERL-AOMDV) was utilized in [14] for transferring data packets in MANET.

Quality of service-enabled ant colony-based multipath routing (QAMR) algorithm was applied in [15] depends on the foraging behavior of ant colony for choosing a path and transmitting data. A maximally radio-disjoint geographic multipath routing protocol (RD-GMR) was designed in [16] to avoid the interference among the multiple paths in MANET.

Fault-Tolerant Disjoint Multipath Distance Vector Routing Algorithm (FD-AOMDV) was developed in [17] that enhance path discovery performance with a reduced amount of delay. A novel multipath routing algorithm was presented in [18] to find out the distinct paths between source and destination nodes with assists of Omnidirectional antennas.

An Energy Reduction Multipath Routing Protocol was utilized in [19] for MANET using Recoil Technique (AOMDV-ER) to preserves the energy along with optimal network lifetime. Trust-based Multi-Path Routing was accomplished in [20] for increasing data security in MANET.

## 6. CONCLUSION

An efficient WT-MRD mechanism is designed with the goal of enhancing performances of multipath routing in MANET with a higher security level. The goal of WT-MRD mechanism is obtained with the application of Neighbor Node-based Trust Calculation (NN-TC) Model and Winnow Linear Multiplicative Classification (WLMC) Algorithm on the contrary to existing works. The proposed WT-MRD mechanism enhances the ratio of a number of mobile nodes that are exactly identified as normal or malicious with a lower amount of time consumption with the assist of NN-TC Model and WLMC Algorithms as compared to state-of-the-art works. Furthermore, the proposed WT-MRD mechanism gets enhanced security by increasing a ratio number of data packets successfully received to the destination from the source node in MANET as compared to conventional works. In addition to that, the proposed WT-MRD mechanism reduces delay time while carried outing the multipath routing in MANET as compared to existing works. The performance of WT-MRD mechanism is determined in terms of attack detection rate, attack detection time and data security level and delay and compared with two existing works. The experimental result illustrates that WT-MRD mechanism provides better performance with an improvement of data security level and minimization of delay when compared to state-of-the-art works.